\documentclass{article}
\usepackage{amsmath,amssymb,bm,color,mathrsfs,graphicx}
\usepackage{authblk}
\usepackage[left=3cm]{geometry}

\newcommand{\B}{\mathscr{B}}
\newcommand{\Bref}{\mathscr{B}_0}
\newcommand{\Brel}{\mathscr{B}_\text{rel}}

\newcommand{\bBe}{\mathbf{B}_\text{e}}

\newcommand{\bBeo}{\mathbf{\mathring{B}}_{\mathrm{e}}}

\newcommand{\bF}{\mathbf{F}}
\newcommand{\bFe}{\mathbf{F}_\text{e}}
\newcommand{\bFeo}{\mathbf{\mathring{F}}_\text{e}}
\newcommand{\bFo}{\mathbf{\mathring{F}}}
\newcommand{\bG}{\mathbf{G}}
\newcommand{\bH}{\mathbf{H}}
\newcommand{\bI}{\mathbf{I}}
\newcommand{\bL}{\mathbf{L}}

\newcommand{\bQ}{\mathbf{Q}}

\newcommand{\bT}{\mathbf{T}}
\newcommand{\bW}{\mathbf{W}}
\newcommand{\bWc}{\mathbf{\hat{W}}}
\newcommand{\sigmaE}{\sigma_\text{e}}
\newcommand{\sigmaF}{\sigma_\text{Fr}}
\newcommand{\sO}{\sigma_\text{iso}}
\newcommand{\bX}{\mathbf{X}}
\newcommand{\ba}{\mathbf{a}}
\newcommand{\bb}{\mathbf{b}}
\newcommand{\be}{\mathbf{e}}

\newcommand{\bh}{\mathbf{h}}
\newcommand{\bk}{\mathbf{k}}
\newcommand{\bkc}{\mathbf{\hat{k}}}
\newcommand{\bn}{\mathbf{n}}
\newcommand{\bt}{\mathbf{t}}
\newcommand{\bu}{\mathbf{u}}
\newcommand{\bv}{\mathbf{v}}
\newcommand{\vs}{v_\text{s}}
\newcommand{\bx}{\mathbf{x}}
\newcommand{\bnu}{\bm{\nu}}
\newcommand{\bomega}{\bm{\omega}}
\newcommand{\bGamma}{\bm{\Gamma}}

\newcommand{\bnp}{\bn_0^{\perp}}
\newcommand{\dotn}{{\dot{\mathbf{n}}}}
\newcommand{\bzero}{\mathbf{0}}
\newcommand{\bna}{\boldsymbol{\nabla}}
\newcommand{\tsp}{\!\top\!}
\newcommand{\F}{\mathscr{F}}
\newcommand{\ph}{\hat{p}}

\newcommand{\D}[2]{\frac{\partial #1}{\partial #2}}

\renewcommand{\phi}{\varphi}
\newcommand{\tp}{\otimes}

\newcommand{\divr}{\mathop{\mathrm{div}}}

\newcommand{\tr}{\mathop{\mathrm{tr}}}

\newcommand{\dev}{\mathop{\mathrm{dev}}}
\newcommand{\piso}{p_{\text{iso}}}

\newcommand{\PH}{\mathbb{P}_{\bH}}
\newcommand{\II}{\mathbb{I}}
\newcommand{\eps}{\epsilon}
\DeclareMathOperator{\re}{Re}

\DeclareMathOperator{\er}{e}

\title{\textbf{Elastic director vibrations \\in nematic liquid crystals}}

\author{Stefano S.\,Turzi%
\thanks{\texttt{stefano.turzi@polimi.it}}}

\affil{\small{Dipartimento\,di\,Matematica, Politecnico\,di\,Milano, Piazza\,Leonardo\,da\,Vinci,\,32~~20133\,Milano\,(Italy)}}

\date{\today}
\begin{document}
\maketitle
\begin{abstract}
Recently Biscari, DiCarlo and Turzi [\emph{Anisotropic wave propagation in nematic liquid crystals}, to appear (arXiv:1311.1802)] proposed a theory for nematoacustics which models nematic liquid crystals as nematic elastomers with molecular relaxation. Here, we extend the analysis of this theory to account for the director motion possibly induced by the propagation of a sound wave. We find that the director vibration is related to the - usually small - anisotropy of the molecular distribution, thus providing a justification to the relative high ultrasonic intensities required to observe non-negligible acousto-optic responses.  
\end{abstract}
%

\section{Introduction}
\label{sec:intro}

The analysis of the interaction of sound waves with the director field still challenge our present understanding of the dynamics of nematic liquid crystals (NLC). For instance, there is no broadly accepted theoretical framework for the description of the sound velocity and the sound attenuation, in particular with respect to their anisotropic features and their frequency or temperature dependence. Other interesting effects, yet not fully understood, include the acousto-optic effect and the acoustic generation. The former is observed when a ultrasonic wave, injected into a NLC cell,  changes the refractive index of the NLC thus altering the optical transmission properties of the cell. Recently, this acousto-optic effect has attracted a renewed interest, due to its potential for application to acoustic imaging \cite{00sand,09sand}. A second manifestation of the coupling between acoustic waves and nematic order is the phenomenon of acoustic generation observed in a NLC cell undergoing Freedericksz transitions 
triggered by an external electric field \cite{99kim}.

Recently, the attempts to a thorough explanation of these phenomena have led to two theories, rather different in nature. The first models the NLC as an anisotropic second-gradient (or Korteweg) fluid \cite{02seli,03seli,04seli,09virga}. Therefore, it puts forward a free-energy density containing a term proportional to $(\bn\cdot\bna \rho)^2$, thus postulating a coupling between the spatial gradient of the mass density, $\bna\rho$, and the nematic director, $\bn$. By contrast, the second is a first-gradient theory characterized by a hyperelastic anisotropic response from an evolving relaxed configuration \cite{14bidi} and the free energy is adapted from the standard theory of nematic elastomers \cite{BTW94,WT96,WT03}. In this paper, we adopt the perspective of the latter theory. 

In the example studied in \cite{14bidi} a number of simplifying assumptions were introduced. In particular, the director field was supposed fixed by a suitable external action and viscosities, other than that associated with the relaxation mechanism, were neglected. Here, we relax one of these assumptions and extend the analysis to include the director vibrations induced by the propagation of the sound wave. An analogous study for the second gradient theory can be found in \cite{11virga}.

This paper is organized as follows. Sec.~\ref{sec:I_background} reviews the theoretical background. Namely, the Cauchy stress tensor, the molecular field and the appropriate balance laws are discussed. Furthermore, the key concept of relaxed configuration and the equation governing its evolution are introduced. The liquid crystal is presented as an anisotropic neo-Hookean elastic material where the relaxed configuration is allowed to evolve. Intuitively, the relaxed configuration is dragged by the present configuration of the fluid, with a characteristic relaxation time which is usually considered to be fast with respect to all the other relaxation times. The following Sec.~\ref{sec:II_wave_propagation} deals with the perturbation analysis of the governing equations in the linear acoustic approximation. Here, the main features of the director vibrations induced by the injected sound wave are derived. Sec.~\ref{sec:III_discussion} contains a discussion, where the subtle interplay of the director motion with 
the 
sound wave is further analysed. Some computational details, possibly useful to the reader, are reported in Appendix.

\section{Theoretical background}
\label{sec:I_background}
In this section, we review the theoretical paradigm developed by Biscari, DiCarlo and Turzi in Ref.\cite{14bidi}, suitably extended to incorporate the director motion. In essence, this theory models nematic liquid crystals as relaxing nematic elastomers, i.e., materials characterized by an anisotropic neo-Hookean elastic energy and a fast molecular rearrangement. This microscopic reorganization, while not affecting the macroscopic deformation, drives the liquid to lower elastic energy states. Thus, in addition to the reference configuration, $\Bref$, and the actual configuration, $\B$, a relaxed configuration, $\Brel$, is introduced which evolves and where the shear stress vanishes. Intuitively, the relaxed configuration is dragged by the actual configuration of the fluid, with a relaxation time which is usually considered to be fast with respect to all the other characteristic times in the system. In doing so, the theory is reconciled with the expectation that a liquid crystal is able to flow.

\begin{figure}[t]
\centering
\includegraphics[width=0.5\textwidth]{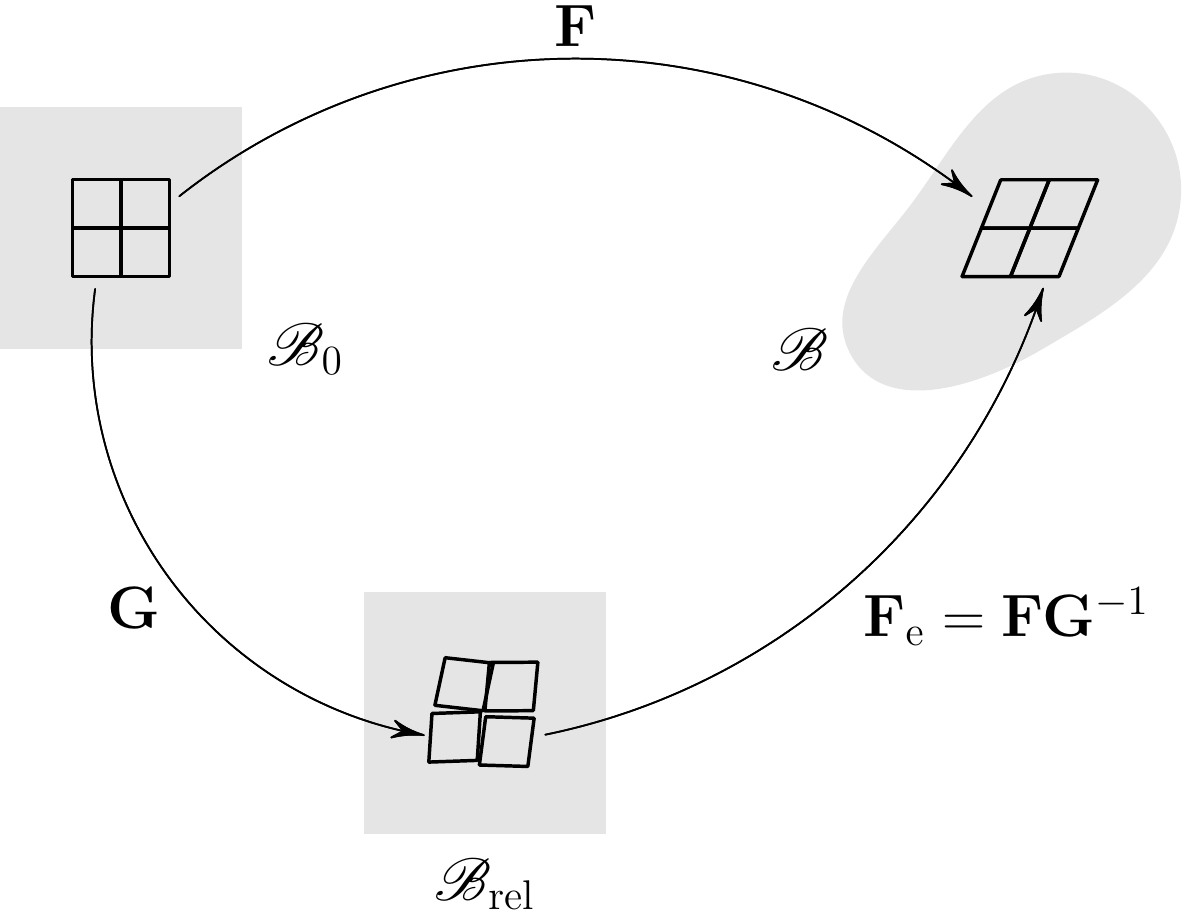}
\caption{Decomposition of the deformation gradient in its accommodating and elastic parts.}
\label{fig:patatoidi}
\end{figure}

In mathematical terms the molecular reorganization can be described through an evolution equation for the relaxed, or natural, configuration $\Brel$ of the nematic liquid crystal. There is a vast literature which deals with this accommodating evolution. For our present purposes we will here mainly refer to the inelastic evolution theory proposed by DiCarlo and co-workers \cite{02dicqui,05adc,07desdicter} or Rajagopal and Srinivasa \cite{98rajaI,98rajaII}. As indicated in Fig.\ref{fig:patatoidi}, we can measure the deformation from the relaxed configuration through the effective tensor $\bFe$, while the relaxing deformation from the reference configuration is described by the tensor $\bG$, so that a multiplicative decomposition holds: $\bF=\bFe \bG$. We remark that the map $\bG$ may be, but it is not requested to be, the gradient of some virtual displacement field, as different material elements may undergo a non-compatible evolution.
The elastic energy stored in any single material element does not explicitly depend on the deformation gradient $\bF$, as this latter measures how much the distorted configuration departs from the original reference configuration $\Bref$, which in general does not coincide with the relaxed configuration. We assume instead that the elastic energy depends on $\bFe=\bF\bG^{-1}$, which measures the distortion of the present configuration with respect to the relaxed configuration. Moreover, frame invariance implies that the free energy may depend only on the effective left Cauchy-Green strain tensor
\begin{equation}
\bBe=\bFe \bFe^{\tsp} = \bF \bH \bF^{\tsp} ,
\label{eq:bBe}
\end{equation}
where we have defined the inverse relaxing right Cauchy-Green strain tensor
\begin{equation}
\bH = (\bG^{\tsp} \bG)^{-1} . 
\end{equation}
Experimental evidence suggests that not all the strains may be recovered by simply reorganizing the natural configuration. In particular, fluids are not able to relax density variations, as each fluid possesses a reference density dictated by the microscopic fact that each molecule occupies in average a well defined volume. As a consequence, a necessary feature of a physically meaningful model of an accommodating fluid is that the energy cost of any density variation should not be compensated by the microscopic relaxation. Therefore, we impose the constraint $\det\bH=1$ to the relaxation strain tensor and we define the isochoric (i.e., unit determinant) tensors as:
\begin{subequations}
\begin{align}
\bFo & = J^{-1/3} \bF, \\
\bFeo & = J^{-1/3} \bFe , \\
\bBeo & = \bFeo \bFeo^{\tsp} = \bFo\bH\bFo^{\tsp}\, ,
\end{align}
\end{subequations}
where $J=\det\bF=\det\bFe$.

\subsection{Elastic energy}
\label{sec:sub_free_energy}
Liquid crystals have a natural preferred microscopic direction, i.e., the director $\bn$. It is then natural to expect that the distribution of the centres of mass of the molecules is affected by the director. This has also been confirmed since the very first realistic Monte-Carlo simulations of the nematic-isotropic-smectic phase transition \cite{93bez}. More precisely, in the presence of elongated molecules, the average distance of the centres of mass of the molecules in a nematic phase depends on the angle between the direction in which it is measured and the director. We assume that the equilibrium distribution of the centres of mass can be described to first approximation by a symmetric second-order tensor $\bL(\rho,\bn)$, resembling the \emph{shape tensor} usually introduced in nematic elastomers \cite{BTW94,WT96,09dester,WT03}. Since we aim at modelling a slightly compressible fluid, the shape tensor $\bL$ is assumed to depend explicitly on the density $\rho$, an 
hypothesis that will turn out to be fundamental. More precisely, we posit
\begin{equation}
\bL(\rho,\bn) = a(\rho)^2 \bn\tp\bn + a(\rho)^{-1}(\bI-\bn\tp\bn) \, ,
\label{eq:Lshape}
\end{equation}
where the tensor product of two vectors, $\ba$ and $\bb$, is the double tensor $\ba\tp\bb$ such that $(\ba \tp \bb) \bu = (\bb\cdot\bu) \ba$ for all $\bu$ (i.e., $(\ba \tp \bb)_{ij} = a_i b_j$). The material parameter $a(\rho)$ provides a quantitative measure of how the onset of nematic order induces an anisotropy in the radial correlation function. The shape tensor is spherical, prolate or oblate respectively for $a(\rho)=1$, $a(\rho)>1$ or $a(\rho)<1$.

In the following we consider a free energy functional which, in addition to the usual Frank potential $\sigmaF(\rho,\bn,\bna\bn)$, includes an elastic term $\sigmaE(\rho,\bBeo,\bn)$. We neglect for simplicity the effect of external actions, as well as any dissipation effect. Furthermore, $\sigmaE(\rho,\bBeo,\bn)$ comprises two terms, intended to penalize two different types of distortions. Alike in all fluids, perturbations of the local average density of molecules involve significant energy variations in NLCs. In view of the \emph{quasi}-incompressible character of nematic liquid crystals it is to be expected that this energy contribution will be the most significant. Energy may be lost/gained if the deformation induces a stretching/compression of the local available area in the direction orthogonal to the director. Therefore, even volume-preserving deformations induce energy variations if they, for example, locally squeeze the material in the directions orthogonal to $\bn$, while 
suitably elongating it in the direction of the director. Under such assumptions the free energy functional in a spatial domain $\B$ is given by
\begin{equation}
\begin{split}
\F := \int_{\B} \big({\tfrac{1}{2}}\rho \bv^2 \!+\! \rho \sigmaE(\rho,\bBeo,\bn) 
\!+\! \rho \sigmaF(\rho,\bn,\bna\bn) \big) dV
\label{eq:Fintegrale}
\end{split}
\end{equation}
where the \emph{strain energy density} per unit mass is 
\begin{subequations}
\begin{equation}
\sigmaE(\rho,\bBeo,\bn) := \sO(\rho) + \tfrac{1}{2}\mu(\rho) \tr\big(\bL(\rho,\bn)^{-1}\bBeo - \bI\big) ,
\label{eq:sigmaE}
\end{equation}
and we use the one-constant approximation for the \emph{Frank potential}
\begin{equation}
\sigmaF(\rho,\bn,\bna\bn) = \tfrac{1}{2} K(\rho)|\bna \bn|^2 .
\label{eq:sigmaF}
\end{equation}
\label{eq:sigma}
\end{subequations}
In the following, we refer to $\sigma = \sigmaE + \sigmaF$ as the \emph{elastic energy density}, per unit mass.

The isotropic contribution $\sO(\rho)$ to \eqref{eq:sigmaE} penalizes density (and volume) variations and is assumed to dominate the elastic energy. The second term in \eqref{eq:sigmaE} couples the elastic properties of the material with the nematic director. Eq.\eqref{eq:sigmaE} mimics the classic elastic energy of nematic elastomers \cite{BTW94,WT96,09dester,WT03} but it differs under two distinguishing features. First, it implements the constitutive prescription that the nematic liquid crystal is able to relax the strain-elastic terms, proportional to the shear modulus $\mu$, through suitable evolution of the relaxed configuration, as this term depend only on the \emph{effective} strain tensor $\bBe$. Secondly, the shape tensor depends explicitly on the density, contrariwise to the standard theories of nematic elastomers, even in the compressible case. 

Finally, it is worth noticing that the equilibrium configuration is achieved by the minimum of the elastic energy, namely
\begin{equation}
\rho=\rho_0, \quad \bv=0, \quad \bn=\text{const.}, \quad  \bBeo = \bL \, .
\end{equation}

\subsection{Balance equations}
\label{sec:sub_balance_equations}
As usually assumed in the continuum description of nematic liquid crystals, the motion of the system is governed by the balance of momentum and the balance of angular momentum at each point of the system. In the absence of body forces and body couples and neglecting the rotational inertia of the director, these two equations are (see \cite{95dgpr,04stewart} and Appendix \ref{app:genfor})
\begin{align}
\rho \dot{\bv} & = \divr \bT \qquad \text{and}\qquad \bn\times\bh=\bzero .
\label{eq:balances}
\end{align}
Here, a superimposed dot indicates the material derivative, $\times$ denotes the cross product, $\bv$ is the macroscopic velocity of the fluid, $\bT$ is the Cauchy stress tensor and $\bh$ is usually called the molecular field. In a hyperelastic material $\bT$ and $\bh$ can be derived by the elastic energy density $\sigma=\sigmaE + \sigmaF$ using the Rayleigh method, suitably extended to continuous bodies \cite{01sovi,04sovi}. Within this method, the material time derivative of $\sigma$ is calculated and $\bT$, $\bh$ are then identified with the generalized forces conjugated to the velocity fields $\bna\bv$ and $\dotn$, respectively. For the interested reader, more details on this derivation are reported in Appendix \ref{app:genfor}. Since $\sigmaE$ only depends on $\rho$ and $\bFeo$, and in view of the identity 
\begin{equation}
\D{\sigma}{\bFo}\bFo^{\tsp} = \D{\sigma}{\bFeo}\bFeo^{\tsp}, 
\end{equation}
we finally obtain
\begin{subequations}
\begin{align}
\bT & =-\rho^2\D{\sigma}{\rho} + \rho \dev \Big(\D{\sigma}{\bFeo}\bFeo^{\tsp}\Big) - \rho(\bna\bn)^{\tsp}\D{\sigma}{\bna\bn}, 
\label{eq:TCauchy_1}\\
\bh & = \rho \D{\sigma}{\bn}-\divr\Big(\rho\D{\sigma}{\bna\bn} \Big),
\label{eq:h_1}
\end{align}
\label{eq:Tandh_1}
\end{subequations}
where $\dev$ is the deviatoric projector: $\dev\bL:=\bL-\tfrac{1}{3}\tr(\bL)\bI$. The substitution of \eqref{eq:sigma} into \eqref{eq:Tandh_1} then yields 
\begin{subequations}
\begin{align}
\bT & =-\,\ph \,\bI + \rho\,\mu(\rho)\dev\!\big(\bL(\rho,\bn)^{-1}\bBeo\big) 
- \rho K(\rho)(\bna\bn)^{\tsp}(\bna\bn), \label{eq:TCauchy_2} \\
\bh & = \rho \mu(\rho) \big(a(\rho)^{-2}\!\!-a(\rho)\big)\bBeo\bn - \divr\big(\rho K(\rho) \bna\bn\big),
\label{eq:h_2}
\end{align}
\label{eq:Tandh_2}
\end{subequations}
where the pressure-like function is
\begin{equation}
\begin{split}
\ph & = \rho^2 \Big(\D{\sO}{\rho} + \tfrac{1}{2}\mu'(\rho) \tr\big(\bL(\rho,\bn)^{-1}\bBeo - \bI\big) \\
& - \tfrac{3}{2}\mu(\rho)\frac{a'(\rho)}{a(\rho)}\dev(\bn\tp\bn)\cdot (\bL(\rho,\bn)^{-1}\bBeo)
+ \tfrac{1}{2}K'(\rho)|\bna\bn|^2 \Big) .
\end{split}
\end{equation}

\subsection{Evolution of the relaxed configuration}
\label{sec:sub_evolution_equation}

Depending on the level of detail we wish to introduce in our model, the microscopic evolution may be characterized by a single or a set of relaxation times. In any case, perfect fluid behaviour is recovered when accommodation is much faster than macroscopic dynamics. By contrast, asymptotically pure elastic behaviour is obtained when the macroscopic dynamics is so fast that the system is in practice not able to accommodate in time. A complete set of dynamic equations requires an explicit evolution equation for the tensor $\bH$. Since $\bH$ belongs to the manifold $\mathcal{M}$ of double symmetric tensors with unit determinant, this evolution is constrained to $\mathcal{M}$. Furthermore, it has to be consistent with the dissipation principle establishing that the power dissipated -- defined as the difference between the power expended and the time derivative of the free energy -- should be non-negative, for all body-parts, at all times. This latter condition localizes into \cite{04raj}:
\begin{equation}
\bT \cdot \bna\bv-\rho\dot{\sigma} \ge 0 .
\label{eq:diss}
\end{equation}
The use of Eqs.\eqref{eq:balances},\eqref{eq:Tandh_1} then simplify this inequality into
\begin{equation}
\frac{\partial\sigmaE}{\partial\bH} \cdot \dot{\bH} \le 0\,.
\label{eq:rediss}
\end{equation}
The simplest possible way to fulfil the requirement \eqref{eq:rediss} is to assume a viscous-like dynamics for the inverse relaxing strain, described by a gradient-flow equation. However, the flow must preserve the symmetry of $\bH$ and the constraint $\det\bH=1$. All this conditions are satisfied, as described in Ref.\cite{14bidi}, if we prescribe the following evolution equation
\begin{equation}
\gamma\,\dot{\bH}=-\,\PH\D{\sigmaE}{\bH}\;.
\label{eq:steep}
\end{equation}
The scalar coefficient $\gamma\!>\!0$ acts as a viscosity modulus, and $\PH\!$ is the orthogonal projector from the space of double tensors onto the subspace tangent to $\mathcal{M}$ at $\bH$:
\begin{equation}
\PH\!=\mathbb{I} - \frac{\bH^{-1}\tp\,\bH^{-1}}{\|\bH^{-1}\|^2}\,,
\label{eq:PH}
\end{equation}
with $\II$ the fourth-order identity tensor. 

On account of \eqref{eq:sigmaE} and \eqref{eq:bBe}, we obtain explicitly
\begin{equation}
\D{\sigmaE}{\bH} = \tfrac{1}{2}\mu(\rho)\bFo^{\tsp}\bL(\rho,\bn)^{-1}\bFo
\label{eq:diffexpl}
\end{equation}
which, substituted into \eqref{eq:steep}, yields
\begin{equation}
\gamma\dot{\bH} - \tfrac{1}{2}\mu(\rho)\big(\bFo^{\tsp}\bL(\rho,\bn)^{-1}\bFo \cdot \bH^{-1}\big)
\frac{\bH^{-1}}{\|\bH^{-1}\|^2} = -\tfrac{1}{2}\mu(\rho) \bFo^{\tsp}\bL(\rho,\bn)^{-1}\bFo \, .
\label{eq:steep_explicit}
\end{equation}

\section{Sound wave and director motion}
\label{sec:II_wave_propagation}
We now examine the propagation of a sound wave by studying the fluid motion and the director vibrations which arise from a slight perturbation of the homogeneous equilibrium configuration. In contrast to Ref.\cite{14bidi}, where the director was held fixed by a suitable external action, we  allow the director to vibrate according to the angular momentum equation \eqref{eq:balances}$_2$. It is to be noticed that, by virtue of the peculiar structure of acoustic perturbations, it is to be expected (and it is indeed the case in second-gradient theories \cite{11virga}) that the induced vibration of the nematic director is a higher order effect. This is also confirmed by experiments, where high acoustic intensities are required to observe an acousto-optic response of the liquid crystal \cite{02seli,03seli,04seli}. As we shall see, this can be interpreted in our model as an indication that the shape tensor at equilibrium is close to the identity tensor, i.e., the distribution of the centres of mass is nearly 
spherical. 

\subsection{Nearly incompressible and slightly anisotropic fluid}
\label{sec:subII_simplifying}
We need to be more specific on the constitutive assumptions related to the functions $\sO(\rho)$ and $a(\rho)$ if we want to describe the effects of a propagating density wave on the director field. To this end, in agreement with Ref.\cite{14bidi}, we introduce the scaled density variation and the isotropic pressure as follows
\begin{equation}
\xi = (\rho-\rho_0)/\rho_0,  \qquad \piso(\rho) = -\rho^2 \D{\sO}{\rho}\, .
\end{equation}
We then posit
\begin{align}
\piso\big(\rho_0(1+\xi)\big) & = p_0 + \rho_0 p_1 \xi + o(\xi) \, , \\
\mu\big(\rho_0(1+\xi)\big) & = \mu_0 + \mu_1 \xi + o(\xi) \, .
\end{align}
The assumption $\xi \ll 1$ accounts for the nearly incompressibility of the liquid crystal, while small anisotropy implies that the shear modulus $\rho_0 \mu_0$ is much smaller than the bulk modulus $\rho_0 p_1$. Therefore, we introduce a small parameter $\eta$ as follows
\begin{equation}
\eta:=\frac{\mu_0}{p_1} \ll 1 \, . 
\end{equation}
Furthermore, we formalize the hypothesis that the anisotropic aspect ratio $a(\rho)$ differs slightly from 1 by positing
\begin{equation}
a(\rho)=a\big(\rho_0(1+\xi)\big)=1+\alpha_0 + \alpha_1 \xi + o(\xi)
\label{eq:aaratio}
\end{equation}
and assuming a small ``asphericity factor'': $|\alpha_0|\ll 1$. In contrast to $\alpha_0$, the \emph{sensitivity coefficient} \mbox{$\alpha_1=\rho_0 a'(\rho_0)$} is \emph{not} required to be small.

\subsection{Perturbation analysis}
The equilibrium configuration, minimum of the free energy, is characterized by a zero velocity field ($\bv_0=0$), a density $\rho_0$, a uniform director field $\bn_0$, i.e., such that $\bna\bn_0 \equiv 0$, and an effective strain tensor \mbox{$(\bBeo)_0=\bL(\rho_0,\bn_0)$}. On account of Eq.\eqref{eq:aaratio} the equilibrium shape tensor and effective strain tensor $\bBeo$ are
\begin{equation}
(\bBeo)_0=\bL(\rho_0,\bn_0) = \bI + 3\alpha_0 \dev(\bn_0\tp\bn_0) + o(\alpha_0),
\end{equation}
which, if we retain only the dominant $O(1)$ order in $\alpha_0$, leads to an equilibrium deformation gradient $\bF= \bI+O(\alpha_0)$. Therefore, within this approximation, we can perturb the equilibrium configuration just identified by simply considering motions of the type
\begin{subequations}
\begin{align}
\bx(\bX,t) &= \bX + \eps \,\ba \re(\er^{i \phi(\bX,t)}), \label{eq:perturabtion_x}\\
\bn(\bx,t) &= \bn_0 + \eps \,\bn_1(\bx,t) \, , \label{eq:perturabtion_n}
\end{align}
\end{subequations}
where $\bX$ is the position vector of a point in the reference configuration, $\phi(\bX,t) = \bkc \cdot \bX - \omega t$ is a complex phase, $\bkc=\bk_r + i \bk_i$ is a complex wave vector, $\eps\ll 1$ is a dimensionless small parameter that scales the amplitude of the wave, and the vector $\ba$ determines the amplitude and the character of the wave (e.g., longitudinal/transversal if $\ba$ is parallel/orthogonal to $\bk_r$). Since $|\bn|=1$, the first order correction to the director field must satisfy $\bn_1 \cdot \bn_0=0$. For later convenience in the calculations, we will retain the complex notation with the implicit understanding that only the real part of the equations has a physical meaning.

The linearised balance law \eqref{eq:balances}$_1$ to order $O(\eps)$ yield the anisotropic sound velocity and attenuation. As we shall see, the director oscillations only contribute to the higher order corrections to the anisotropic velocity and attenuation. Hence, the results reported in Ref.\cite{14bidi} continue to hold to leading order also in the presence of director motion. However, we are here mainly concerned with a complementary problem: we want to study to what extent the director field is influenced by the macroscopic velocity field of the fluid. In this case, the governing equation is obtained from the $O(\eps)$ approximation of the torque balance \eqref{eq:balances}$_2$.

Neglecting $O(\alpha_0)$ terms, and to order $O(\eps)$, Eqs.\eqref{eq:perturabtion_x},\eqref{eq:Lshape} and \eqref{eq:aaratio} yield
\begin{align}
\dot{\bv} &= -\eps \rho_0 \omega^2 \er^{i\phi}\ba , \\
\bF &= \bI+i\eps\er^{i\phi}(\ba \tp \bkc), \\
\det(\bF) & =1 + i\eps\er^{i\phi}(\ba \cdot \bkc), \\
\xi & = -i \eps \er^{i\phi} (\ba \cdot \bkc), \\
\bFo & = \bI+i\eps\er^{i\phi}\dev(\ba \tp \bkc), \\
\bL & = \bI + 3\alpha_1 \xi \dev(\bn_0 \tp \bn_0) \, .
\end{align}
After writing $\bH=\bI + \eps \bH_1$, the evolution equation \eqref{eq:steep_explicit} can be linearised to give
\begin{equation}
\tau \dot{\bH}_1 + \bH_1 = -i\er^{i\phi}\dev\big(\ba\tp\bkc+\bkc\tp\ba + 3 \alpha_1 (\ba\cdot\bkc) (\bn_0\tp\bn_0) \big)
\end{equation}
with solution (modulo an exponentially decay transient)
\begin{equation}
\bH = \bI - i\eps \frac{\er^{i\phi}}{1-i\omega\tau} \dev\big(\ba\tp\bkc+\bkc\tp\ba + 3 \alpha_1 (\ba\cdot\bkc) (\bn_0\tp\bn_0) \big),
\end{equation}
where $\tau=2\gamma/\mu_0$. This allow us to calculate the isochoric effective strain tensor
\begin{equation}
\bBeo = \bI + i\eps \er^{i\phi} \dev\Big(\frac{-i\omega\tau}{1-i\omega\tau}(\ba\tp\bkc+\bkc\tp\ba) 
- \frac{3\alpha_1}{1-i\omega\tau} (\ba\cdot\bkc)(\bn_0\tp\bn_0)\Big) ,
\label{eq:Beo}
\end{equation}
and its product with the inverse shape tensor
\begin{align}
\bL(\rho,\bn)^{-1}\bBeo = \bI\! +\! i\eps\!\er^{i\phi}\!\!\frac{-i\omega\tau}{1-i\omega\tau} 
\dev\big((\ba\tp\bkc+\bkc\tp\ba) + 3\alpha_1 (\ba\cdot\bkc)(\bn_0\tp\bn_0)\big) . 
\label{eq:LmBeo} 
\end{align}
One remarkable consequence of Eq.\eqref{eq:LmBeo} is that the first order perturbation to the director field, i.e. $\eps\bn_1$, does not enter the calculation of the Cauchy stress tensor to the considered order of approximation. Therefore, the balance law \eqref{eq:balances}$_1$, which leads to the anisotropic speed of sound and attenuation, is not affected by the small oscillations of the director field as these yield only second order corrections. Thus, the $O(\eta)$ solution of the balance of linear momentum \eqref{eq:balances}$_1$, where $\eta:=\mu_0/p_1$, is found to be in the form of a longitudinal plane wave with anisotropic velocity given by
\begin{align}
\frac{\vs}{\sqrt{p_1}} & = 1+\eta f(\omega\tau)\big(\tfrac{2}{3}\!-\!\alpha_1\!+\!\tfrac{3}{2}\alpha_1^2+3\alpha_1 (\cos\theta)^2\big), 
\label{eq:vsAnisotropic}
\end{align}
where $\vs$ is the speed of sound, $\theta$ is the angle between the propagation direction $\be$ and the uniform director field $\bn_0$, and the function $f(x)$ is 
\begin{equation}
f(x) = \frac{x^2}{1+x^2} \quad (x\ge 0). 
\end{equation}
The attenuation vector then reads
\begin{equation}
\sqrt{p_1}\,\bk_i = \frac{\eta}{\tau} f(\omega\tau)\Big(\big(\tfrac{2}{3}\!-\!\alpha_1\!+\!\tfrac{3}{2}\alpha_1^2
+3\alpha_1 (\cos\theta)^2\big)\be + \tfrac{3}{2}\alpha_1 (\sin2\theta) \bt \Big), 
\label{eq:attAnisotropic}
\end{equation}
where $\bt$ is a unit vector orthogonal to $\be$ in the plane $\text{span}\{\be,\bn_0\}$ and such that $\bn_0\cdot\bt>0$.
We refer the reader to Ref.\cite{14bidi} for a more detailed derivation of these results and for the comparison of Eqs.\eqref{eq:vsAnisotropic},\eqref{eq:attAnisotropic} with the experimental data reported in \cite{70lord,72mull}.

We now turn the attention to the balance of torques \eqref{eq:balances}$_2$ in order to derive the leading approximation to the director motion. After some algebra which we omit for brevity, we find
\begin{align}
& \bBeo\bn \times \bn = i\eps \er^{i\phi} \frac{-i\omega\tau}{1-i\omega\tau} 
\big((\bkc\cdot\bn_0)(\ba\times\bn_0) + (\ba\cdot\bn_0)(\bkc\times\bn_0) \big) , \\
& \divr\big(\rho K(\rho) \bna \bn\big)\times \bn = \eps \rho_0 K(\rho_0) (\Delta \bn_1) \times \bn_0
\end{align}
So far, the validity of our approximation has been limited to zeroth order in $\alpha_0$ and to first order in $\eps$. However, in view of the fact that $\bBeo\bn \times \bn = O(\eps)$ and that 
\begin{equation}
a(\rho)^{-2}-a(\rho) \sim -3(\alpha_0 + \alpha_1 \xi) = -3\alpha_0 + 3 i \eps \er^{i\phi} \alpha_1 (\ba \cdot \bkc) ,
\label{eq:a(rho)coeff}
\end{equation}
we gather that the torque balance \eqref{eq:balances}$_2$, with $\bh$ as given in \eqref{eq:h_2}, is valid also to first order in $\alpha_0$. In fact, the $O(\alpha_0)$ corrections to $\bBeo$ do not contribute to this equation as $\bBeo\bn \times \bn$ is to be multiplied by \eqref{eq:a(rho)coeff}, which is a small coefficient in our approximation. Therefore, it is apposite to keep terms up to $O(\alpha_0)$ in the $O(\eps)$ approximation to the \eqref{eq:balances}$_2$ which then reads
\begin{align}
K_0 (\Delta \bn_1) & \times \bn_0 + 3\alpha_0 \mu_0 i\er^{i\phi}\frac{-i\omega\tau}{1-i\omega\tau} 
\big((\bkc\cdot\bn_0)(\ba\times\bn_0) + (\ba\cdot\bn_0)(\bkc\times\bn_0) \big) = 0,
\label{eq:directorEq1}
\end{align}
where $K_0=K(\rho_0)$. We now look for oscillating solutions of \eqref{eq:directorEq1} and posit
\begin{equation}
\bn = \bn_0 + \eps\bn_1 = \bn_0 + \eps\re(\er^{i\phi}\bWc) \bn_0 , 
\label{eq:assumptionbn}
\end{equation}
where $\bWc=\bW_r + i \bW_i$ is a complex matrix and $\bW_r$, $\bW_i$ are skew-symmetric matrices to account for the constraint $\bn_1 \cdot \bn_0=0$. This implies that
\begin{align}
\bn_1 & = \er^{i\phi} \bWc \bn_0 \\
\Delta \bn_1 & = -(\bkc\cdot\bkc) \er^{i\phi} \bWc \bn_0 , 
\end{align}
where, consistently with the previous notation, we implicitly consider only the real part of this expressions. Furthermore, the results of Ref.\cite{14bidi} show that at leading order the sound wave is purely longitudinal, with vanishing imaginary wave vector. Thus, we posit
\begin{equation}
\bk_r = \frac{2\pi}{\lambda_0} \be, \qquad \bk_i = 0, \qquad \ba = a_0 \be , 
\end{equation}
where $\lambda_0$, $a_0$ are respectively the wavelength and the amplitude of the sound wave, and $\be$ is the unit vector which identifies the propagation direction. Eq.\eqref{eq:directorEq1} then simplifies to
\begin{equation}
\bWc\bn_0 \times \bn_0 = \frac{3}{\pi}\alpha_0 \frac{\lambda_0}{\delta_0} 
\frac{\omega\tau}{1-i\omega\tau}(\be\cdot\bn_0)(\be\times\bn_0),
\label{eq:directorEq2}
\end{equation}
where we have introduced the correlation length $\delta_0$, defined as
\begin{equation}
\delta_0 = \frac{K_0}{\mu_0 a_0},
\end{equation}
which is to be compared to the wavelength $\lambda_0$. When $\lambda_0 \ll \delta_0$ the director distortions induced by the sound wave are contrasted by the high energy cost associated to the Frank potential. By separating the real and the imaginary parts of \eqref{eq:directorEq2}, the following solutions are found
\begin{subequations}
\begin{align}
\bW_r\bn_0 &= \frac{3}{\pi}\alpha_0 \frac{\lambda_0}{\delta_0} 
\frac{\omega\tau}{1+(\omega\tau)^2} \sin(2\theta) \, \bnp ,  \\
\bW_i\bn_0 &= \frac{3}{\pi}\alpha_0 \frac{\lambda_0}{\delta_0} 
\frac{(\omega\tau)^2}{1+(\omega\tau)^2} \sin(2\theta) \, \bnp ,
\end{align}
\label{eq:directorEq3}
\end{subequations}
where $\bnp$ is the unit vector in the plane $\text{span}\{\be,\bn_0\}$ orthogonal to $\bn_0$ such that the propagation direction $\be$ is decomposed as: $\be = \cos\theta \,\bn_0 + \sin\theta \,\bnp$. We recall that, due to the assumption \eqref{eq:assumptionbn}, $\bW_r\bn_0$ and $\bW_i\bn_0$ have to be multiplied by $\cos\phi$ and $\sin\phi$, respectively, to yield the director oscillation $\bn_1$. Therefore, after some algebra, we can convert these two orthogonal oscillations into an amplitude and phase representation to obtain
\begin{equation}
\bn_1 =  \frac{3}{2\pi}\alpha_0 \frac{\lambda_0}{\delta_0}\frac{\omega\tau}{\sqrt{1+(\omega\tau)^2}} \sin(2\theta) \, \bnp \cos\big(\bk\cdot\bx-\omega\tau + \beta(\omega\tau)\big),
\label{eq:amplitude_n1}
\end{equation}
where $\beta(\omega\tau)=\arctan\omega\tau$ measures the phase delay of the director oscillation with respect to the sound wave.

\section{Discussion}
\label{sec:III_discussion}
Nematic liquid crystals consist of organic molecules that interact weakly; as a result, molecular order is easily perturbed and quite modest fields or boundary effects are sufficient to cause a quite massive reorganization and to influence strongly their structure and macroscopic properties. As such, external actions, such as magnetic or electric field, generally modifies greatly the director field. By contrast, the dynamics of defects in NLCs is often studied under the \emph{no-backflow} approximation where the director rotation induced by the presence of a macroscopic flow (and vice versa) is neglected. Thus, the interaction of the director with the underlying fluid is seen as a higher order effect.

When a sound wave propagates in a NLC cell, it induces an undulatory macroscopic motion that interacts with the orientational order of the NLC. Given the above remarks on the no-backflow approximation, it is maybe not surprising that the director field is not perturbed to leading order by the fluid motion. This is especially true because we have explicitly neglected all the viscosities except that associated to the evolution of the relaxed configuration. Therefore, the anisotropic speed of sound \eqref{eq:vsAnisotropic} and the wave attenuation \eqref{eq:attAnisotropic}, as  derived in \cite{14bidi} under the simplifying assumption that the director is held fixed by an external action, continue to hold to first order also when the director is free to vibrate.

Strictly speaking these considerations apply when the asphericity factor $\alpha_0$ is vanishing small. However, when $\alpha_0$ is small but not negligibly small, we can push the perturbation analysis further to include the $O(\alpha_0)$ corrections. An immediate consequence of \eqref{eq:amplitude_n1} is that the director oscillation amplitude is proportional to $\alpha_0$, thus providing an indication that an anisotropic molecular distribution, i.e., an anisotropic equilibrium shape tensor, could be responsible for more sensitive acousto-optic responses. As far as the angular dependence is concerned, \eqref{eq:amplitude_n1} shows that the amplitude is maximum when $\theta = \pi/4$, while it vanishes for either the propagation direction, $\be$, parallel or perpendicular to the director.

More subtle is the discussion of the frequency dependence. If the perturbation to the equilibrium configuration is slow enough, the fluid is able to relax its natural configuration so to avoid storing any elastic energy. This corresponds to an ``instantaneous'' attainment of the equilibrium configuration $\bL^{-1}\bBeo=\bI$ which, in the small-frequency limit, leads to vanishing anisotropic effects. Mathematically, this limit is achieved by fixing $\omega$ (in order to fix also $\lambda_0$) and let $\tau \to 0$ such that $\omega\tau \ll 1$. We gather from \eqref{eq:amplitude_n1} that also the director oscillations disappear in this limit. On the other extreme, when the wave period is much shorter than the relaxation time ($\omega\tau \to +\infty$), the nematic liquid crystal behaves effectively as an anisotropic elastic solid, as it has no time to rearrange its natural configuration. In this limit, the anisotropic effects on the sound speed, the attenuation and the director motion become frequency-
independent and saturate to a maximum value. It is to be remarked that the fact that the attenuation saturates to a maximum value in the high-frequency regime is a consequence of the approximation in which we have neglected the usual nematic viscosities, whose effect would have resulted in an unbounded attenuation.

\section*{Acknowledgements} The author would particularly like to thank Paolo Biscari and Antonio DiCarlo for their encouragement and for many enlightening discussions. The kind  support of the Italian Ministry of University and Research through the Grant No.\,200959L72B\underline{\;\;}004 `Mathematics and Mechanics of Biological Assemblies and Soft Tissues' is also acknowledged.

\bibliographystyle{siam}
\bibliography{references}

\appendix

\section{Generalized forces in nematic liquid crystals}
\label{app:genfor}
Here, we show explicitly how to identify the generalized forces in a nematic liquid crystal and thus determine the formulas for the Cauchy stress tensor and the molecular field as given in Eqs.\eqref{eq:TCauchy_1}, \eqref{eq:h_1}. To this end, we can resort to the method described in \cite{01sovi,04sovi}. For simplicity we neglect the effect of external actions, as well as any dissipation effect, and the kinetic contribution which stems from the molecular rotation. Under such assumptions, the free energy functional in a spatial domain $\B$ is given by
\begin{align}
\F & = \int_{\B} \Big({\textstyle{\frac{1}{2}}}\rho \bv^2 + \rho \sigma(\rho,\bFo,\bn,\bna\bn)\Big) dV
\end{align}
with $\sigma = \sigmaE(\rho,\bFo,\bn) + \sigmaF(\rho,\bn,\bna\bn)$. We now compute the time-derivative of the free energy in order to identify the generalized forces. By making use of Reynolds' Transport Theorem and with the aid of the identity \mbox{$(\bna\bn )^{\bm{\cdot}} =  \bna \dotn - (\bna\bn) \bna \bv$}, we obtain
\begin{equation}
\dot{\F} =\int_{\B} \rho \bv \cdot \dot{\bv}
+ \rho \Big(\D{\sigma}{\bF}\cdot\dot{\bF}+\D{\sigma}{\bn}\cdot \dotn + \D{\sigma}{\rho}\dot{\rho}
+ \D{\sigma}{\bn}\cdot \dotn + \D{\sigma}{\bna\bn}\cdot (\bna \dotn - (\bna\bn) \bna \bv) \Big) dV .
\label{eq:F_punto}
\end{equation}
Let us now compute separately the different terms in Eq.\eqref{eq:F_punto}:
\begin{align}
\rho\D{\sigma}{\bF} \cdot\dot\bF & = \rho\D{\sigma}{\bF}\bF^{\tsp} \cdot \bna\bv
= \dev\Big(\rho\D{\sigma}{\bFo}\bFo^{\tsp}\Big) \cdot\bna\bv, \\
\rho \D{\sigma}{\rho}\dot{\rho} & = -\rho^2 \D{\sigma}{\rho} \,\bI\cdot \bna\bv , \label{eq:app_sempA4} \\
\rho\D{\sigma}{\bna\bn}& \cdot (\bna \bn)\bna\bv =\rho\,(\bna \bn)^{\tsp}\D{\sigma}{\bna\bn}\cdot \bna\bv,
\end{align}
where, in \eqref{eq:app_sempA4}, we have used the mass balance equation: 
\begin{equation}
\dot{\rho}+\rho\divr \bv = 0 . 
\end{equation}
We can further simplify Eq.\eqref{eq:F_punto} if we consider the following identities
\begin{align}
\rho\D{\sigma}{\bna\bn}\cdot \bna \dotn & = \divr \big(\bGamma^{\tsp}\dotn\big) - (\divr\bGamma)\cdot \dotn, \\
\bQ\cdot \bna \bv & = \divr \big(\bQ^{\tsp}\bv \big) - (\divr\bQ)\cdot \bv,
\end{align}
where $\bQ$ is an arbitrary tensor and we have introduced the couple tensor \mbox{$\bGamma=\rho(\partial\sigma/\partial\bna\bn)$}. If we now define the Cauchy stress tensor $\bT$ and the molecular field $\bh$ as in Eq.\eqref{eq:Tandh_1}, we find
\begin{equation}
\dot{\F} = \int_{\B} \Big[(\rho \dot{\bv}-\divr\bT)\cdot \bv +\bh\cdot \dotn \Big]\, dV 
+ \int_{\partial\B} \Big[\bT\bnu\cdot\bv+\bGamma\bnu\cdot \dotn\Big]\, da,
\label{eq:F_punto_2}
\end{equation}
where use has been made of the Divergence Theorem, and $\bnu$ denotes the outer unit normal to $\partial\B$. We finally remark that, since $\bn$ is a unit vector, its time derivative $\dotn$ is orthogonal to $\bn$. In fact, a vector $\bomega$ can be defined such that
$\dotn = \bomega \times \bn$. We can therefore rewrite Eq.~\eqref{eq:F_punto_2} as
\begin{equation}
\dot{\F} = \int_{\B} \Big[(\rho \dot{\bv}-\divr\bT)\cdot \bv +\bn\times\bh\cdot\bomega\Big]\, dV 
+ \int_{\partial\B} \Big[\bT\bnu\cdot\bv+\bn\times\bGamma\bnu\cdot \bomega\Big]\, da.
\label{eq:F_punto_3}
\end{equation}

The use of the Rayleigh method \cite{01sovi,04sovi} and the observation that $\bv$ and $\bomega$ in \eqref{eq:F_punto_3} may attain arbitrary values leads to the evolution equations of hyperelastic nematic liquid crystals, which (in the absence of viscous forces, body forces and body couples) are
\begin{align}
\rho \dot{\bv} & = \divr \bT \qquad \text{and}\qquad \bn\times\bh=\bzero.
\end{align}
Analogously, it emerges that the stress $\bT\bnu$ and the torque $\bn\times\bGamma\bnu$ are to be balanced by external actions applied on $\partial\B$.
\end{document}